\documentclass{article}
\usepackage{ijcai21}

\usepackage{times}

\usepackage{soul}
\usepackage{url}
\usepackage[hidelinks]{hyperref}
\usepackage[utf8]{inputenc}
\usepackage[small]{caption}
\usepackage{graphicx}
\usepackage{amsmath}
\usepackage{booktabs}

\usepackage{amsfonts}
\usepackage{subfigure}
\usepackage{multirow}
\usepackage{color}
\usepackage{algorithm}  
\usepackage{algpseudocode}  

\usepackage{diagbox}
\newcommand\Algphase[1]{%
\vspace*{-.4\baselineskip}\Statex\hspace*{\dimexpr-\algorithmicindent-2pt\relax}\rule{0.473\textwidth}{0.4pt}%
\Statex\hspace*{-\algorithmicindent}\textbf{#1}%
\vspace*{-.7\baselineskip}\Statex\hspace*{\dimexpr-\algorithmicindent-2pt\relax}\rule{0.473\textwidth}{0.4pt}%
}

\title{T-WaveNet$:$ Tree-Structured Wavelet Neural Network for Sensor-Based \\Time Series Analysis}

\author{
Minhao Liu$^1$\footnote{Contact Author}\and
Ailing Zeng$^1$\and
Qiuxia Lai$^1$\And
Qiang Xu$^1$\\
\affiliations
$^1$The Chinese University of Hong Kong\\
\emails
\{mhliu, alzeng, qxlai, qxu\}@cse.cuhk.edu.hk
}

\begin{document}

\maketitle

\begin{abstract}
Sensor-based time series analysis is an essential task for applications such as activity recognition and brain-computer interface. Recently, features extracted with deep neural networks (DNNs) are shown to be more effective than conventional hand-crafted ones. However, most of these solutions rely solely on the network to extract application-specific information carried in the sensor data. Motivated by the fact that usually a small subset of the frequency components carries the primary information for sensor data, we propose a novel tree-structured wavelet neural network for sensor data analysis, namely \emph{T-WaveNet}. To be specific, with T-WaveNet, we first conduct a power spectrum analysis for the sensor data and decompose the input signal into various frequency subbands accordingly. Then, we construct a tree-structured network, and each node on the tree (corresponding to a frequency subband) is built with an invertible neural network (INN) based wavelet transform. By doing so, T-WaveNet 
provides more effective representation for sensor information than existing DNN-based techniques, and it achieves state-of-the-art performance on various sensor datasets, including UCI-HAR for activity recognition, OPPORTUNITY for gesture recognition, BCICIV2a for intention recognition, and NinaPro DB1 for muscular movement recognition.
\end{abstract}

\section{Introduction}
Sensors play an essential role in our everyday life. For example, inertial sensors in the smartwatch are used for human activity logging~\cite{chen2020deep}. Electroencephalography (EEG) signals for brain intention classification can help disabled people communicate effectively~\cite{zhang2020motor}. Surface electromyography signals are widely used in muscular movement analysis for clinical diagnosis and prosthetic device control~\cite{ahmad2009surface}.


Generally speaking, sensor-based time series analysis consists of three steps: (i). \emph{data segmentation}, wherein the continuous sensor data is partitioned into segments using fixed- or variable-sized windows; (ii). \emph{feature extraction}, wherein various techniques are applied on each segment to extract discriminative features.; and (iii). \emph{downstream tasks}, which perform the required tasks (e.g., classification or prediction) with the extracted features. No doubt to say, feature extraction is the most critical step among them.

Sensor features can be broadly categorized as hand-crafted features and deep learning-based features. The former includes temporal features (e.g., the mean and the variance of the sensor signal) and frequency features obtained by performing signal processing techniques such as discrete wavelet transform (DWT), mean frequency (MNF), and power spectrum ratio (PSR). Temporal features are simple to calculate~\cite{janidarmian2017comprehensive,qian2019novel}, but they are not effective for complicated tasks. By contrast, frequency features are more informative by preserving the context information in different frequency components~\cite{atzori2014electromyography,phinyomark2013emg,duan2015semg}, but the feature extraction process is often laborious and time-consuming.  Recently, deep neural networks (DNNs) have become the dominant approaches for feature extraction in various fields, and DNN-based features also show superiority for sensor-based time-series analysis, e.g.,~\cite{rahimian2020xceptiontime,ignatov2018real,yao2017deepsense}.

Existing DNN-based solutions typically extract features from the raw sensor signals directly without considering different frequency components' roles. In practice, a small subset of the frequency components always carries the primary information for sensor data, known as the \textit{dominant energy range} of the signal~\cite{rioul1991wavelets}. 
Motivated by it, we take signal energy ranges into consideration and propose a tree-structured wavelet neural network, namely \emph{T-WaveNet}, to extract more effective features. 
The main contributions of this work are threefold:

\begin{itemize}


\item We perform \emph{power spectrum analysis} to decompose the input signals into multiple frequency subbands and utilize them to configure the tree structure in \textit{T-WaveNet}. Such flexible design is applicable to various types of sensor data. 



\item The tree node in T-WaveNet is realized as a novel invertible neural network (INN) based wavelet transform unit, which can effectively separate the input signal's frequency subbands. To the best of our knowledge, this is the first attempt to model the wavelet transform using INN.

\item Finally, we propose an instrumental \emph{feature fusion module} according to the feature dependencies across different frequency components to deal with the personalized heterogeneity from the sensor signal.

\end{itemize}


With the above, T-WaveNet enhances task-related subband feature extraction, thereby dramatically improving the performance of downstream tasks. Extensive experiments on four real-world sensor-based recognition tasks, namely activity recognition, gesture recognition, intention recognition, and muscular movement recognition, show that T-WaveNet consistently outperforms the state-of-the-art solutions by a considerable margin.

In Section 2, we give a brief overview of related works. Section 3 details our proposed \textit{T-WaveNet}. Next, we introduce performance metrics and our experimental settings in Section 4 and compare \textit{T-WaveNet} with state-of-the-art methods and conduct ablation study in Section 5. Finally, Section 6 concludes this work.

\begin{figure*}[htb]
\centering
\includegraphics[width=160mm]{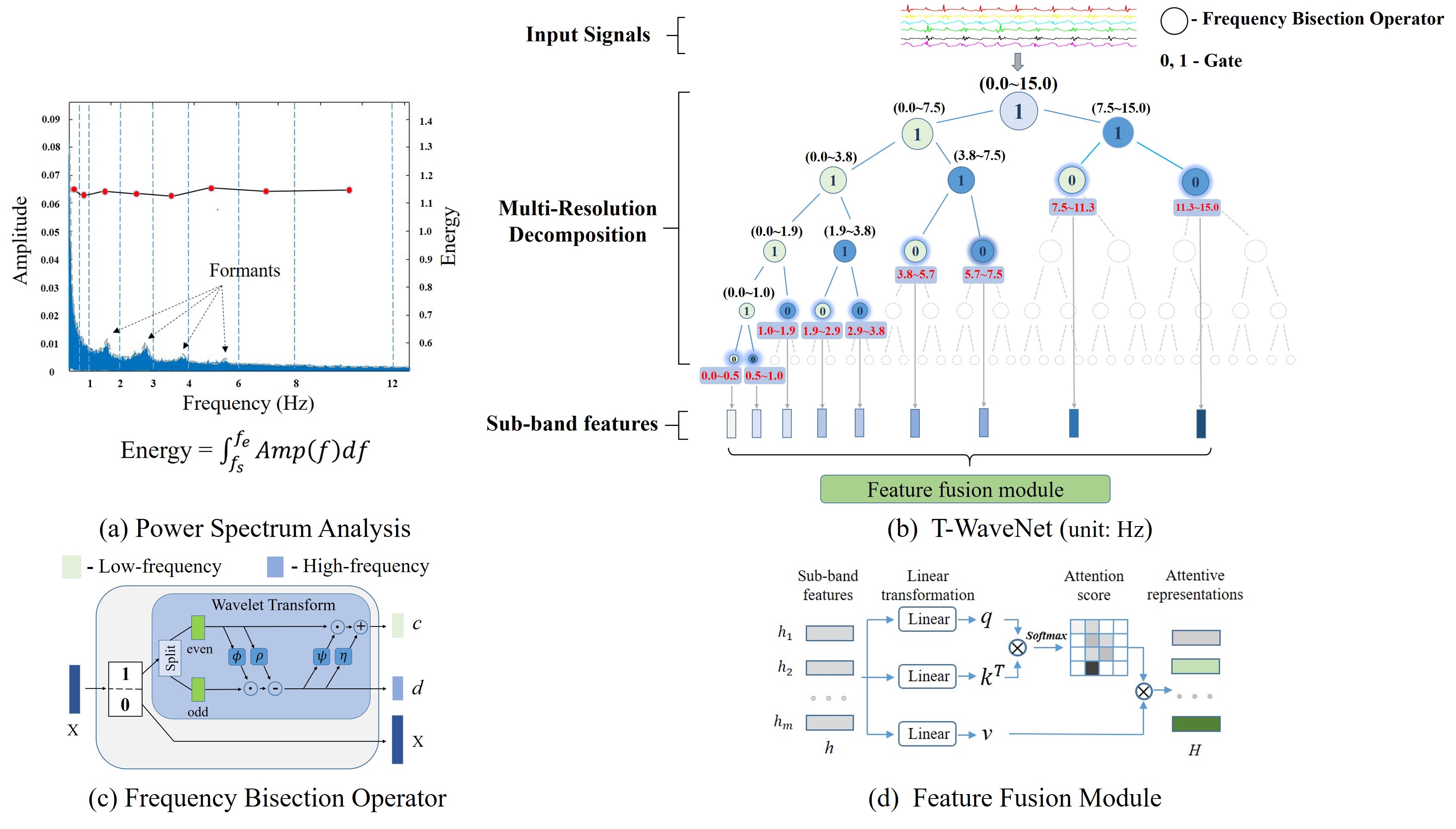}
\caption{
Given an input signal, we first perform \textbf{(a) Power Spectrum Analysis} to decompose it into multiple frequency subbands with comparable energy, where each subband contains at most one formant. Then, based on these frequency subbands, we construct  \textbf{(b) T-WaveNet}, a tree-structured network, where each node is a  \textbf{(c) Frequency Bisection Operator} built with an INN-based wavelet transform unit. The operator outputs the high- and low-frequency components of the input if its binary gate value is ``1", and otherwise bypass the input. Considering the personalized heterogeneity of the input, we utilize a \textbf{(d) Feature Fusion Module} to fuse the subband features $\{h_i\}$ according to the feature dependencies across various frequency components. 
}
\label{fig:diagram}
\vspace{-10pt}
\end{figure*}

\section{Related Work}
Sensor-based time-series analysis is a common task and has been studied in applications such as human activity recognition \cite{geng2016gesture}, brain activity identification \cite{lawhern2018eegnet}, heart rate monitoring \cite{labati2019deep}, and muscular assessment \cite{wei2019surface}.
Our proposed T-WaveNet model advances the development of sensor-based time series analysis by addressing two key challenges: 
(1) Feature extraction of time-series sensor data, 
and (2) Effective modeling of wavelet transform.
\subsection{Feature Extraction of Sensor Data}
Sensor features can be broadly classified into hand-crafted features and deep learning-based features, and the former can be further divided into temporal and frequency features.

Temporal features (e.g., mean, variance, and root mean square (RMS)) are widely used in early studies~\cite{vepakomma2015wristocracy,janidarmian2017comprehensive,qian2019novel}. Though simple to calculate, these temporal features are not effective for complicated tasks. 
Frequency features are more informative than temporal features because they preserve the context information in different frequency components, and have shown improved performances various tasks. \cite{duan2015semg} performs discrete wavelet transform on surface-Electromyography (sEMG) signal for hand motion classification. \cite{geng2016gesture,wei2019multi,jiang2015human,li2017motion} applies the Short-time Discrete Fourier Transform (STDFT) to extract features for human activity recognition. However, this is often a labor-intensive and time-consuming process~\cite{qian2019novel,wang2019deep}. 

Recently, deep learning-based features have shown superiority in various tasks in both performance and efficiency.
For example, \cite{zeng2014convolutional,yang2015deep,ignatov2018real,zeng2020,lee2017human,xing2019driver} consider using convolutional neural networks (CNNs) to extract feature along time dimension. \cite{guan2017ensembles,yao2017deepsense,chen2016lstm,ordonez2016deep} combine CNNs with Recurrent neural networks (RNNs) to extract local and long-term temporal dependency features. Later, \cite{zhang2020motor} integrates the graph neural network to represent the positioning information of EEG nodes, which achieves superior performance. 

However, the existing methods usually extract features from the raw sensor data, and do not distinguish the roles of various frequency components. In contrast, our proposed framework not only utilizes deep learning techniques for efficient feature extraction, but alleviates the above problems by constructing a tree-structured network with the input of multiple split frequency subbands to differentiate the fused components in sensor data analysis.

\section{Proposed T-WaveNet Solution}
Instead of solely relying on the network structure for information extraction from sensor data, our T-WaveNet utilize the knowledge within the data itself to construct a task-specific tree-structured network for feature learning. 

An overview of the \emph{T-WaveNet} is shown in Fig.~\ref{fig:diagram}. 
We first perform a power spectrum analysis on the sensor data (Fig.~\ref{fig:diagram}(a)) to decompose it into multiple frequency subbands with comparable energy, where each subband contains at most one formant. Then, based on these frequency subbands, we construct a tree-structured network (Fig.~\ref{fig:diagram}(b)), where each node is a frequency bisection operator (Fig.~\ref{fig:diagram}(c)) built with an INN-based wavelet transform unit. The operator is conditioned on a binary gate. It outputs the high-frequency and low-frequency components of the input if the gate value is ``1", and otherwise bypass the input. 
Additionally, since different users have their unique style, resulting in the diversity of signal patterns. Thus, to deal with such personalized heterogeneity problem, after obtaining the set of leaf features for each frequency subband, we utilize an effective feature fusion method which enhances the task-related subband features with larger fusion weights (Fig.~\ref{fig:diagram}(d)). 

\subsection{Power Spectrum Analysis}

Considering that only a small subset of the frequency components carries the primary information for sensor data, 
we first perform power spectrum analysis to obtain the set of frequency subbands that covers the dominant energy range of the signal, where each subband contains approximately equivalent energy. Then, we construct T-WaveNet, the tree-structured network wherein each of the leaf nodes corresponds to a frequency subband. 

Alg.~\ref{alg:Framework} shows the power spectrum analysis consisting of two phases: a \emph{formant-guided frequency band splitting} phase to obtain the initial subbands set, and an \emph{energy-guided advanced subband splitting} to further balance the energy of each subband. In the first phase, given the input signal $X\in\mathbb{R}^N$, we first calculate the spectrum with Fourier transform (Fig.~\ref{fig:diagram}(a)). Next, we obtain the set of formants $\mathcal{P}=\{f_p\}_p$, where $f_p$ is a local maximum of the envelope of spectrum. Such formants represent the most direct source of the signal information. Then, we recursively bisecting the frequency band until there is at most one formant falls in each frequency subband $[f_s, f_e]$, where $f_s$ and $f_e$ are the starting and ending frequency, respectively. All these subbands are collected in set $\mathcal{Q}$.
In the second phase, we calculate the energy of each subband in $\mathcal{Q}$ with Eq.(1), and recursively bisecting the subband whose energy exceeds twice the minimum subband energy $E_{\text{min}}$, which is to ensure each subband contains a relatively same amount of information. Because we argue that \emph{evenly-distributed information would enhance network representation ability and reduce the burden of the learning process}. 
$$
Energy = \int_{f_s}^{f_e}Amp\left ( f \right )d_f     \eqno{(1)}
$$
where $Amp(f)$ is the amplitude of the frequency $f$.

\begin{algorithm}[!t]
\caption{Power Spectrum Analysis}\label{alg:Framework}
\begin{algorithmic}[1]
\Algphase{Phase 1 - Formants guided frequency band splitting}
\Require Input signal $X$, frequency subband set $\mathcal{Q}=\emptyset$
\Ensure $\mathcal{Q}$ (updated)
\Procedure{Itv\_Bisect}{$[f_s, f_e], \mathcal{P}, \mathcal{Q}$}
\State $f_m = (f_s+f_e)/2$
\If{Count$(f_p\!\in\![f_s, f_m])\!>\!1$ for $f_p\!\in\!\mathcal{P}$}
    \State \Call{Itv\_Bisect}{$[f_s, f_m], \mathcal{P}, \mathcal{Q}$}
\Else
    \State $\mathcal{Q}\gets\mathcal{Q}\cup \{[f_s, f_m]\}$
\EndIf
\If{Count$(f_p\!\in\![f_m, f_e])\!>\!1$ for $f_p\!\in\!\mathcal{P}$}
    \State \Call{Itv\_Bisect}{$[f_m, f_e], \mathcal{P}, \mathcal{Q}$}
\Else
    \State $\mathcal{Q}\gets\mathcal{Q}\cup \{[f_m, f_e]\}$
\EndIf
\State \textbf{return} $\mathcal{Q}$
\EndProcedure%
\State $Amp(f)\gets FFT(X), f\!\in\![0, F]$ \Comment{Perform Fourier transform on $X$ to get the power spectrum}
\State Obtain the set of formants $\mathcal{P}=\{f_p\}_p$ \Comment{The local maximum of the envelope of $Amp(f)$.}
\State \Call{Itv\_Bisect}{$[0, F],\mathcal{P},\mathcal{Q}$}


\Algphase{Phase 2 - Energy guided advanced subband splitting}
\Require $\mathcal{Q}=\{[f_i, f_{i+1}]\}_i$, $Amp(f)$
\Ensure $\mathcal{Q}$ (updated)
\Procedure{E\_Bisect}{$[f_i, f_{i+1}], E_i, E_{\text{min}},\mathcal{Q},Amp(f)$}
\If{$E_i\!>\!2*E_{\text{min}}$}
\State $f_m = (f_i+f_{i+1})/2$
\State $\mathcal{Q}\gets \mathcal{Q}/\{[f_i, f_{i+1}]\}$ 
\State $\mathcal{Q}\gets\mathcal{Q}\cup \{[f_i, f_m], [f_m, f_{i+1}]\}$
\State $E_{i1}\gets \int_{f_i}^{f_m} Amp(f)df$ 
\State $E_{i2}\gets\int_{f_m}^{f_{i+1}}Amp(f)df$
\State \Call{E\_Bisect}{$[f_i, f_m], E_{i1}, E_{\text{min}},\mathcal{Q}$}
\State \Call{E\_Bisect}{$[f_m, f_{i+1}], E_{i2}, E_{\text{min}},\mathcal{Q}$}
\EndIf
\State \textbf{return} $\mathcal{Q}$
\EndProcedure
\For{$[f_i, f_{i+1}]\!\in\!\mathcal{Q}$}
\State $E_i\gets\int_{f_i}^{f_{i+1}} Amp(f)df$ 
\EndFor
\State $\mathcal{E}\equiv\{E_i\}, E_{\text{min}}\gets \text{min }\mathcal{E}$
\For{$E_i\!\in\!\mathcal{E}$}
\State \Call{E\_Bisect}{$[f_i, f_{i+1}], E_i, E_{\text{min}},\mathcal{Q}$}
\EndFor
\end{algorithmic}
\end{algorithm}

After obtaining the subband set $\mathcal{Q}$, we construct the tree-structured network of T-WaveNet in three steps: (i). \textit{bottom-up} marking; (ii). \textit{top-down} completion; and (iii). \emph{pruning}. 
In the bottom-up process, given a full binary frequency tree with some height (see the dashed tree structure in Fig.~\ref{fig:diagram}(b)), we locate each subband in $\mathcal{Q}$ and set its binary gate as ``0", which serves as the leaf nodes. Then, we set the binary gates of all other nodes on the path from the leaves to the root as ``1". 
In the top-down process, for all the nodes with gate ``1", we set the gates of their children as ``0" if the gates have not been configured in the previous process. 
Finally, the nodes without gate settings are pruned from the tree. 
The resulted sub-tree of the full binary tree preserves the leaf nodes of the informative frequency subbands in  $\mathcal{Q}$, as well as some closely-related frequency components that are not in $\mathcal{Q}$ (\textit{i.e.}, some nodes added in the top-down process) for robustness and tolerances.

\subsection{Frequency Bisection Operator}
We build our wavelet transform in the frequency bisection operator based on the so-called \emph{Lifting Scheme}~\cite{sweldens1998lifting},
known as the second-generation wavelets. It is a simple and powerful approach to construct different wavelets, e.g., \emph{Haar}~\cite{haar1909theorie}.  The main idea is to utilize the strong correlation among the neighboring samples on the signal to separate the low-frequency (approximation) and high-frequency subband (details), respectively. 
The workflow of the Lifting Scheme contains three stages. 
Assuming the input vector is $X = (x[0],x[1],...,x[N])$.
\begin{itemize}
\item  \textit{Splitting}. The signal is split into two non-overlapping partitions. The general partition method is dividing the signal to even part $X_{even} = (x[0],x[2],...,x[2k])$ and odd part $X_{odd} = (x[1],x[3],...,x[2k-1])$ , $k \in N/2$. The spliting operater is:
$$
(X_{even}, X_{odd}) = Splitting(X)        \eqno{(2)}
$$
\item  \textit{Predictor}. The intersections of two partition sets are distributed in the original signal. Based on the signal correlation, given one of them, it is possible to build a good predictor P for the other set. $d$ means the difference between the given and the prediction set. This step is also called ``dual lifting".
$$
d = X_{odd} - P(X_{even})     \eqno{(3)}
$$
\item  \textit{Updator}. Using the odd part to update the even part with updator U in order to keep some consistent characteristics of the original signal, e.g., mean, higher moments. This step is called "primal lifting".
$$
c = X_{even} + U(d)     \eqno{(4)}
$$
\end{itemize}
However, such wavelet construction framework with fixed coefficients ($P, U$) is suboptimal and restricts the adaptability in DNNs. Therefore, we take advantage of the invertible neural networks (INN), a bijective transformation, which can effectively build correlations between inputs and outputs with learnable structures. Therefore, to adjust the Lifting Scheme to an adaptive wavelet transform, the improvements include: (i). we revise the Eq. $(3),(4)$ to an affine function as Eq. $(5),(6)$, which can enhance the transformation ability; (ii). we realize the coefficients($\phi$, $\psi$,$\rho$ and $\eta$ ) with convolutional layers.
By doing so, our adaptive wavelet transformation can learn the wavelet coefficients from the data directly, which should be more optimal for input signals. 
$$
d = X_{odd}\odot   exp(\phi (X_{even}) ) - \rho(X_{even})     \eqno{(5)}
$$
$$
c = X_{even}\odot exp(\psi (d) ) + \eta (d)     \eqno{(6)}
$$
where $\phi$ and $\psi$ stand for scale, $\rho$ and $\eta$ stand for translation. $\odot$ is the
element-wise production. As shown in Fig.~\ref{fig:diagram}(c). Note that the $exp(\cdot)$ is omitted in Fig.~\ref{fig:diagram}(c) around the function $\phi$ and $\psi$.

\begin{table*}[tph]
  \centering
  \fontsize{8}{9}\selectfont 
  \label{tab:performance_comparison}
    \begin{tabular}{c|c|c|c|c|c|c|c}

    \hline
     Datasets        & Classes             &Window size         &Sensor types & Feature number   &Sampling rate(Hz)  &Subjects\cr\hline
\hline
     OPPORTUNITY     &18                &48 &Inertial Sensors                  &77                       &30              &4\cr\hline
     UCI-HAR         &6                 &128   &Inertial Sensors   &9           &50 &30\cr\hline
     BCICIV2a        &4              &400   &Electrodes   &22                  &250&9\cr\hline
     NinaPro DB1     &52          &150      &Electrodes          &10     &100      &27\cr\hline
    \hline
    \end{tabular}
     \caption{The overall information of the four datasets. “Window size” means the length of the input signal used in the experiments. “Features number" denotes the input channel in each dataset.}
\label{table:dataset}
\end{table*}






\subsection{Adaptive Feature Fusion} 
 Considering the personalized heterogeneity of the sensor data, the feature extracted from different frequency components should have different contributions to a specific task. Therefore, we employ an efficient feature fusion module which is realized by self-attention mechanism~\cite{vaswani2017attention}. The module learns the feature dependencies across various frequency components and enhance the more task-related features by assigning higher weights. 
Following~\cite{vaswani2017attention}, we describe our attention model as a mapping from a query and a set of key-value pairs to an output, as illustrated in Fig.~\ref{fig:diagram}(d). The sub-band features $h \in \mathbb{R}^{c \times m}$ extracted from tree structure are first used to form the query and key-value pairs. 
$$
q = W_{q}h+b_{q},  q\in \mathbb{R}^{d \times m},
$$
$$
k = W_{k}h+b_{k},  k\in \mathbb{R}^{d \times m},   \eqno{(7)}\\  
$$
$$
v = W_{v}h+b_{v},  v\in \mathbb{R}^{d \times m}
$$
Where $W_q, W_k, W_v\in \mathbb{R}^{d \times c}$ and $b_q, b_k, b_v \in \mathbb{R}^{d}$ are the weight matrix and bias. $m$ and $c$,$d$ denote the number of features and feature dimensions, respectively. For each key, we then compute an attention score as 
$$
\alpha = softmax( k^{T} \cdot q), \alpha \in \mathbb{R}^{m \times m}   \eqno{(8)}
$$
Then the output of the attention module is computed by the weighted sum of the attention score and the values.
$$
H =  \alpha \cdot v,   H \in \mathbb{R}^{m \times d}          \eqno{(9)}
$$      
Therefore, the attentive frequency sub-band representation $H$ is fed into a standard softmax classifier:
 $$
        p_i = softmax(WH+b)               \eqno{(10)}
$$
where $W$ and $b$ are the weight matrix and bias, respectively. $p_i$ is the predicted probability for class $i$.

\subsection{Loss Function}
In addition to employing the Cross-Entropy loss function for classification, we also explore the utility of the regularization term to constrain the wavelet decomposition during training. The loss function is shown in Eqn.$(11)$, where $C$ denotes the number of classes, $y_i$ is the binary indicator which equals $1$ if class label $i$ is the correct classification for observation.  The regularization term ensures that for each frequency bisection operator, the mean value of the decomposition output $c_{j}$ is close to that of the input $x_j$ \cite{rodriguez2020deep}. $M$ is the total number of operators in the tree-structured network. $\lambda$  tunes the strength of the regularization term. 
$$
Loss = -\sum_{i}^{C}y_ilog(p_i) + \lambda \sum_{j}^{M}\left \| x_j-c_{j} \right \|_2  \eqno{(11)}
$$

\section{Experimental Setup}



\begin{figure*}[tbh]
\centering  

\subfigure[UCI-HAR ]{
\label{Fig.network.2}
\includegraphics[width=0.32\textwidth]{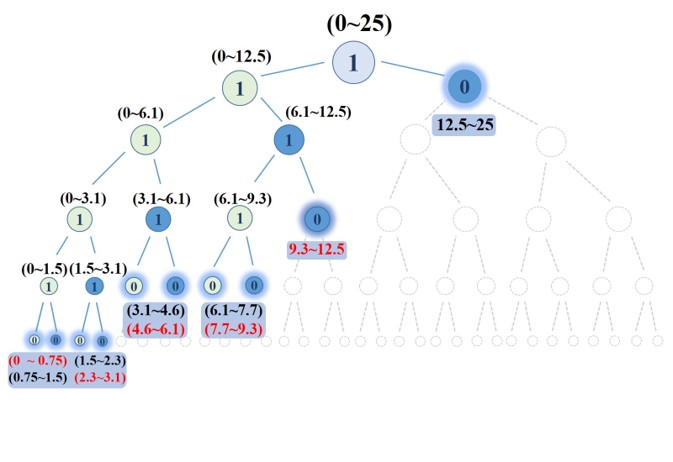}}
\subfigure[BCICIV2a ]{
\label{Fig.network.3}
\includegraphics[width=0.32\textwidth]{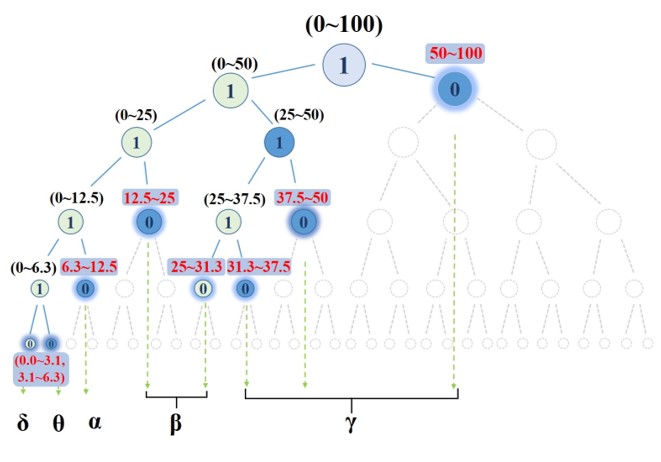}}
\subfigure[NinaPro DB1]{
\label{Fig.network.4}
\includegraphics[width=0.32\textwidth]{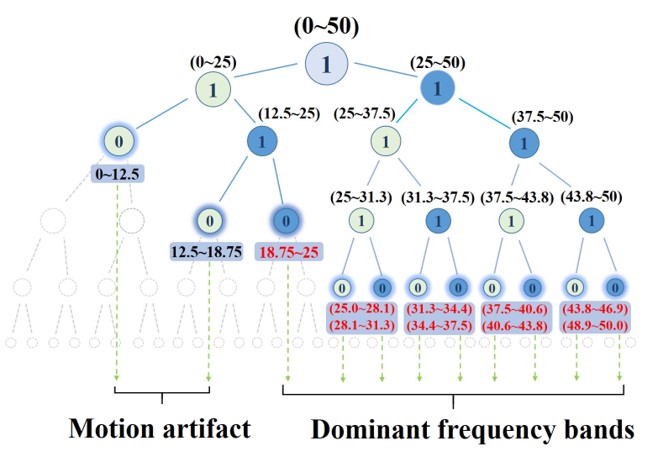}}

\caption{The network structure of different signal types based on the power spectrum analysis. The dominant frequency subbands based on formants are highlighted with red color. (unit:Hz)}
\label{Fig.structure}
\end{figure*}

\subsection{Datasets}
We conduct experiments on four datasets, namely OPPORTUNITY dataset (OPPOR)~\cite{sagha2011benchmarking}, UCI-HAR~\cite{anguita2013public}, the BCICIV2a~\cite{brunner2008bci}, and NinaPro DB1~\cite{atzori2012building}.
The overall statistics of the datasets are listed in Table~\ref{table:dataset}. 
\textbf{OPPOR} consists of annotated recordings from on-body sensors when carrying out common gestures of kitchen activities (\textit{e.g.}, Open Door and Close Door). We follow the settings of the sporadic gestures task in OPPORTUNITY challenge~\cite{chavarriaga2013opportunity}.
\textbf{UCI-HAR} collects sensor data of $6$ activities (walking, walking upstairs, walking downstairs, sitting, standing, laying) from $30$ volunteers aged 19 to 48 years old. We follow the official dataset configuration~\footnote{https://archive.ics.uci.edu}.  
\textbf{BCICIV2a} contains EEG signals from $9$ healthy subjects performing four movement intention tasks (left hand, right hand, feet, and tongue), which are bandpass-filtered between 0.5Hz and 100Hz. We use the same settings as~\cite{zhang2020motor}.
\textbf{NinaPro DB1} contains sparse multi-channel sEMG recordings for hand prostheses, and we configure this dataset following~\cite{rahimian2020xceptiontime}. 
\begin{table*}[tph]
  \centering
  \fontsize{9}{10}\selectfont
  \label{tab:performance_comparison}
    \begin{tabular}{c|c|c|c|c|c|c|c}
    \hline
    \multirow{2}{*}{Methods}&
    \multicolumn{3}{c|}{UCI HAR}&\multicolumn{2}{c|}{OPPOR}&{NinaPro}&{BCICIV2a}\cr\cline{2-8}
    &$F_m$&$F_w$&$Accuracy$&$F_m$&$F_w$&$Accuracy$&$Accuracy$\cr
    \hline
    \hline
     DeepConvLSTM\cite{ordonez2016deep}      &0.9032&0.9054     &0.9080  &0.704 &0.915             &-                  &-\cr
     CNN\cite{ignatov2018real}&-&-&0.9235&-&-&-&-\cr
     D$^2$CL\cite{xi2018deep}           &-&-    &-   &0.7107&0.9197            &-                  &-\cr
    
     Harmonic \cite{hu2020harmonic}     &0.9312&0.9292    & 0.9298 &0.5752&0.8944                  &-                  &-\cr
     FilterNet\cite{chambers2020filternet}              &-&-     & - &0.7430&0.9280                  &-                  &-\cr\hline
     MV-CNN\cite{wei2019surface}            &-&-           & -     &- &-                     &0.8740             &-\cr
     XceptionTime\cite{rahimian2020xceptiontime}      &-&-     &-            &- &-                     &0.9181            &-\cr\hline
     EEgNet\cite{lawhern2018eegnet}            &-&-                 &- &-                &   -  &-                  &0.5130\cr
     NG-CRAM\cite{zhang2020motor}           &-&-                 &- &-                   & - &-                  &0.6011\cr
     \hline
    T-WaveNet-Haar      &0.8918&0.8925       &0.8924           &0.6437   &0.9083           &0.8503                  &0.4312\cr
     T-WaveNet (without feature fusion)   &0.9521&0.9516& 0.9539  &0.7473 &0.9277 &0.9201                   &0.6103\cr
    
     T-WaveNet        &{\bf 0.9642}&{\bf 0.9638}&{\bf0.9705}&{\bf 0.7633}&{\bf 0.9310}&{\bf 0.9321}&{\bf  0.6301}\cr
    \hline
    \end{tabular}
     \caption{Overall comparison results on the four datasets. ``Haar" means the frequency bisection operator is replaced by traditional haar wavelet.}
     \label{table:results}
\end{table*}



\subsection{T-WaveNet Configurations}
Based on the power spectrum analysis, the decomposed frequency subbands and the tree-structured network for each dataset is detailed as follows.
The configuration of OPPOR dataset is shown in Fig~\ref{fig:diagram}(b), which is configured based on the spectrum of Fig.~\ref{fig:diagram}(a), and those for the other three datasets are illustrated in Fig.~\ref{Fig.structure}. As for the power spectrum of UCI HAR, the most dominant frequency range is below 12.5 Hz~\cite{karantonis2006implementation}.
Also, due to the subject-specific motor imagery~\cite{cr2011analysis}, the dominant energy range of BCICIV2a falls in various frequency bands, including $\delta$ ($0.1$-$3$ Hz), $\theta$ ($4$-$7$ Hz), $\alpha$ ($8$-$13$Hz), $\beta$ ($14$-$30$ Hz) and $\gamma$ ($31$-$100$Hz). 
For NinaPro DB1, the frequency bandwidth of sEMG during muscle contraction is $20$-$2000$Hz~\cite{jamaluddin2016filtering} and the motion artifact noise falls mostly in $0$-$20$Hz~\cite{jamal2012signal}. Considering that the sampling rate of NinaPro DB1 is only $100$Hz, we concentrate more on signal within $20$-$50$Hz (\emph{Nyquist}~\cite{grenander1959nyquist}).
\subsection{Network Details} 
\noindent\textbf{Frequency bisection operator.} The $\phi$, $\psi$, $\rho$, and $\eta$ in Eqn. $(4)$ and $(5)$ are modeled as DNN modules. For the T-WaveNet used to obtaining the results in Table~\ref{table:results}, the module structure is: Conv1D($3, 1$)$\rightarrow$LeakyRelu($\alpha=0.01$)$\rightarrow$Dropout($rate=50\%$)$\rightarrow$Conv1D($3, 1$)$\rightarrow$Tanh. The output channel of the first and the second convolutions are three times and one times the input channel, respectively.

\noindent\textbf{Feature fusion module.} The dimension of key/query/value $d$ in Eqn. $(6)$ is $32$. The input channel $c$ is different for each dataset, as listed in Table~\ref{table:dataset}. 



\noindent\textbf{Classifer.} The structure is: FC($32, 1024$) $\rightarrow$ BatchNorm $\rightarrow$ LeakyRelu($\alpha=0.01$) $\rightarrow$FC($ 1024,\#class$)$\rightarrow$Softmax.


\subsection{Training Details} 

All the experiments are run on a single Nvidia GTX 1080 Ti. The batch size is $64$ for all datasets.  We use Adam optimizer~\cite{kingma2014adam} with initial learning rate $3*10^{-4}$ and the decay is $0.95$. The maximum number of training epoch is $100$. $\lambda$ in Eqn. $(11)$ is $0.1$ for all the experiments. The input channel $c$ and the sliding window size of each dataset are listed in Table~\ref{table:dataset}. 

\subsection{Evaluation Metrics}
Following previous works, we use $Accuracy$, $weighted$ $F1$ $score$ $(F_w)$ and $macro(mean)$ $F1$  $score$ $(F_m)$ as our evaluation metrics. The definitions are as follows:
$$
Accuracy = \frac{Number \, of \, correct \, classification}{Total \, number \, of \, test \, samples}   \eqno{(12)}
$$

$$
F_w =  2\sum_{i=0}^{C-1} w_i \frac{precision_i\times recall_i}{precision_i+ recall_i} \eqno{(13)}
$$
$$
F_m = \frac{2}{C} \sum_{i=0}^{C-1}\frac{precision_i\times recall_i}{precision_i+ recall_i} \eqno{(14)}
$$
where $i$ is the class index, $w_i=N_i/\sum_{i=0}^{C-1}N_i$ is the proportion of samples of the class, and $N_i$ is the number of samples in $i$th class. $C$ is the total number of the class. 

\section{Experimental Results}
In this section, we present detailed experimental results to demonstrate the advantages of T-WaveNet design. 

\begin{figure*}[htb]
\centering
\includegraphics[width=140mm]{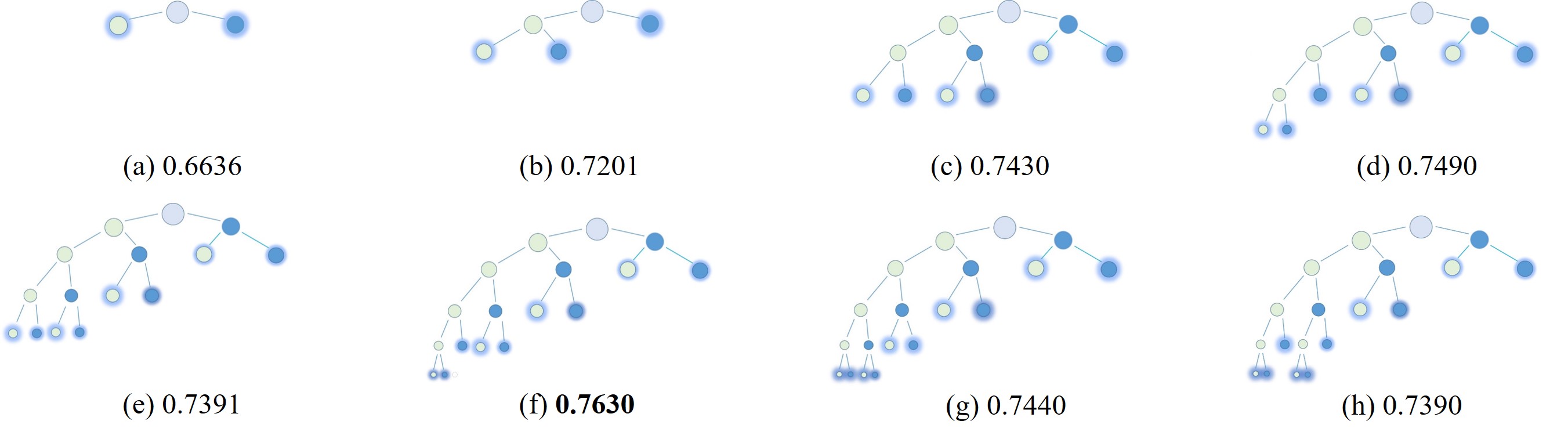}
\caption{The $F_m$ results of different energy divisions in the OPPOR dataset.}
\label{fig:ablation}
\end{figure*}

\subsection{Comparison with the State-of-the-Arts}
To evaluate the overall performance of the proposed model, we compare our model with other methods in each dataset. The results are shown in Table~\ref{table:results} where scores of other methods are mainly from the original papers.  
Our proposed \textit{T-WaveNet} achieves the best performance on all datasets. Specifically, our method yields relative $4.38\%$, $3.72\%$, $3.54\%$ improvements of the $accuracy$, $F_w$ score, $F_m$ score metrics over the $2$nd best method \cite{hu2020harmonic} on UCI-HAR, respectively.
For OPPOR, we improve over the $2$nd best method \cite{chambers2020filternet} by relative $2.66\%$. 
Moreover, we achieve relative $1.52\%$ and $4.81\%$ improvements on NinaPro and BCICIV2a, respectively.

\subsection{Ablation Study}
We study the effectiveness of each proposed components by performing a detailed ablation study on the challenging and popular OPPOR dataset.
\vspace{5pt}
\noindent\textbf{Effectiveness of power spectrum analysis}. 
In Alg.~\ref{alg:Framework}, we first conduct a formant-guided subband splitting process to obtaining the initial subband set $\mathcal{Q}$, and then perform an energy-guided advanced subband splitting. Such subband splitting scheme ensures that each resulted subband contains roughly equivalent energy (information). 
To verify the effectiveness of power spectrum analysis, we train the variants of the T-WaveNet by altering the constraints about the formants and the energy in each decomposed frequency subband, resulting in different levels of tree-structured networks. 
Fig.\ref{fig:ablation} shows the network structures and the corresponding $F_m$ scores. The default tree structure (Fig.\ref{fig:ablation}(f)) outperforms all other variants with either coarser (shallower trees in Fig.\ref{fig:ablation}(a-e)) or finer energy division (deeper trees in Fig.\ref{fig:ablation}(g-h)). Based on the above results, the set of frequency subbands with evenly distributed energy is beneficial for the representation learning of sensor data, and the proposed solution facilitates to find such configuration effectively.

\vspace{5pt}
\noindent\textbf{Impact of frequency bisection operator}.
Our frequency bisection operator is built as an INN-based wavelet transform. 

To verify the impact of such design, we first experiment with the variant, \emph{T-WaveNet-Haar}, where $\phi$, $\psi$, $\rho$ and $\eta$ are traditional Haar wavelet basis. The results in Table~\ref{table:results} shows that our \emph{T-WaveNet} achieves $18.6\%$ improvement on $F_m$ score over the Haar variant. This is because our deep wavelet basis can be learnt from data, thus being more adaptive to various sensor signals compared with Haar wavelet.

Next, we experiment on the deep version of the Lifting Scheme, in which we realize the Predictor $P$ and Updater $U$ in Eqn.$(3)(4)$ using the same deep modules as $\phi$, $\psi$ ,$\rho$ and $\eta$. Besides the default architecture, we also experiment with other structures, as shown in Table~\ref{table:conv}. In the table, L represents the number of convolution layers. C indicates the times of the number of output channel compared with the input channel. D is the dilation size of the first convolution layer. 
As can be seen, our method consistently outperforms the deep Lifting Scheme version across all the network structure settings on the two evaluation metrics, indicating the superiority of the INN-based operator.

\begin{table}[h]
\begin{center}
\fontsize{7}{8}\selectfont
\begin{tabular}{c|c|c|c|c}
\hline
\multirow{2}{*}{Configuration}&
    \multicolumn{2}{c|}{$P, U$}&\multicolumn{2}{c}{$\phi$,$\psi$,$\rho$,$\eta$}\cr\cline{2-5}
                   &$F_m\%$&$F_w\%$&$F_m\%$&$F_w\%$\cr
\hline
default:(L-2, C-3, D-1)&92.6&73.5&\textbf{93.1}&\textbf{76.3}\\
\hline
(L-1, C-3, D-1)&91.2&69.4&\textbf{91.9}&\textbf{71.4}\\
(L-3, C-3, D-1)&91.5&71.6&\textbf{92.4}&\textbf{73.5}\\
\hline
\hline
(L-2, C-1, D-1)&92.7&74.5&\textbf{92.8}&\textbf{75.5}\\
(L-2, C-5, D-1)&91.6&72.1&\textbf{92.7}&\textbf{75.7}\\
(L-2, C-6, D-1)&92.3&73.0&\textbf{92.4}&\textbf{74.6}\\
(L-2, C-10, D-1)&91.9&71.2&\textbf{92.0}&\textbf{72.3}\\
\hline
\hline
(L-2, C-3, D-2)&91.9&71.9&\textbf{92.5}&\textbf{74.4}\\
(L-2, C-3, D-3)&92.0&73.6&\textbf{92.7}&\textbf{75.1}\\
(L-2, C-3, D-4)&92.0&72.9&\textbf{92.7}&\textbf{75.0}\\
(L-2, C-3, D-5)&92.0&72.0&\textbf{92.5}&\textbf{73.3}\\
\hline

\end{tabular}
\caption{
Comparison of Lifting scheme and INN-based wavelet transform units.
L-$k$ means the module has $k$ convolution layers; C-$k$ represents the output channel size of the first layer is $k$ times of the input; D-$k$ is the dilation size of the first convolution layer to enlarge the receptive fields.
}
\label{table:conv}
\end{center}
\vspace{-10pt}
\end{table}




\vspace{5pt}
\noindent\textbf{Feature fusion module}. 
To demonstrate the effectiveness of the proposed adaptive feature fusion module for handling the personalized heterogeneity of the sensor data, we remove the fusion module from T-WaveNet and fuse the leaf features $\{h_i\}$ with equivalent weights. As shown in Table~\ref{table:results}, the performance of the resulted model decrease by $2\%$-$3\%$, which indicates that the proposed feature fusion module can make better classification for sensor data recorded from different subjects, because it fuses the features of various frequency subbands adaptively based on the dependencies among multiple frequency components.  

\section{Conclusion}
We propose \textit{T-WaveNet}, a novel tree-structured wavelet neural network for sensor-based time series analysis. By decomposing the input signal into frequency subbands according to power spectrum analysis and utilizing INN-based frequency bisection operator, T-WaveNet provides more effective representation for sensor information over existing DNN-based feature extraction techniques. Experimental results on four different kinds of sensor datasets show that \textit{T-WaveNet} consistently outperforms the state-of-the-art solutions by a considerable margin.

\bibliographystyle{named}

\end{document}